\documentclass[aps,aps, llncs, preprint]{revtex4}
\usepackage{amsmath}
\usepackage{amsfonts}
\usepackage{amssymb}
\usepackage{graphicx}
\usepackage{amsmath}
\usepackage{float}
\usepackage{color}
\usepackage{soul}
\usepackage{siunitx}
\usepackage{physics}
\usepackage[utf8]{inputenc}
\usepackage[T1]{fontenc}

\DeclareUnicodeCharacter{0096}{_}
\DeclareUnicodeCharacter{2009}{_}

\begin{document}
\title{High transport spin polarization in the van der Waals ferromagnet Fe$_4$GeTe$_2$} 

\author{Deepti Rana$^1$, Monika Bhakar$^1$, Basavaraja G.$^2$, Satyabrata Bera$^3$, Neeraj Saini$^1$, Suman Kalyan Pradhan$^3$, Mintu Mondal$^3$, Mukul Kabir$^2$ and Goutam Sheet$^1$} 

\email{goutam@iisermohali.ac.in}

\affiliation{$^1$Department of Physical Sciences, Indian Institute of Science Education and Research (IISER) Mohali, Sector 81, S. A. S. Nagar, Manauli, PO 140306, India}

\affiliation{$^2$Department of Physics, Indian Institute of Science Education and Research (IISER) Pune, PO 411008, India}

\affiliation{$^3$School of Physical Sciences, Indian Association for the Cultivation Of Science, Jadavpur, Kolkata-700032}

\begin{abstract}
\textbf{The challenging task of scaling-down the size of the power saving electronic devices can be accomplished by exploiting the spin degree of freedom of the conduction electrons in van der Waals (vdW) spintronic architectures built with 2D materials. One of the key components of such a device is a near-room temperature 2D  ferromagnet with good metallicity that can generate a highly spin-polarized electronic transport current. However, most of the known 2D ferromagnets have either a very low temperature ordering, poor conductivity, or low spin polarization. In this context, the Fe$_n$GeTe$_2$ (with $n\geq3$) family of ferromagnets stand out due to their near-room temperature ferromagnetism and good metallicity. We have performed spin-resolved Andreev reflection spectroscopy on Fe$_4$GeTe$_2$ ($T_{Curie} \sim$ 273 K) and demonstrated that the ferromagnet is capable of generating a very high transport spin polarization, exceeding 50$\%$. This makes Fe$_4$GeTe$_2$ a strong candidate for application in all-vdW power-saving spintronic devices.}

\end{abstract}

\maketitle


High electrical conductivity, high ferromagnetic Curie temperature ($T_{Curie}$), large spin polarization at the Fermi level of a material and its ease of integrating with other materials are some of the most important prerequisites for application in spintronic devices where the spin degree of freedom of the conduction electron is exploited for novel functionalities\cite{Hirohata,Wolf,Zutic,Bader}. While metal–oxide–semiconductor (CMOS)-based spintronic devices\cite{Makarov} have contributed to the society immensely, scaling-down the size of such devices has become increasingly challenging. Without such down-scaling, the increasing demand for power saving data storage and processing cannot be fulfilled. With the emergence of 2D material systems as a viable platform for the realization of complex functional devices through vdW heterostructuring, 2D vdW ferromagnets have emerged as candidates for application in beyond-CMOS, low-power spintronic devices\cite{Ahn, Hu, Liu, Han, Yang, Sierra, Avsar, Lin, Dayen}. In such ferromagnets, long-range ordering is stabilized in vdW- coupled nearly 2D layers that can be easily exfoliated and integrated with other materials. Such ferromagnets also provide the scope of additional functionality due to the intrinsic anisotropy in the magnetic properties. Most interestingly, it is remarkable that the ferromagnetic order is robust down to very few monolayers of the vdW ferromagnets primarily due to strong spin-orbit interaction\cite{Gong, Pinto}. All these features collectively enable the 2D ferromagnets to give rise to unprecedented emergent properties with diverse application potential.

However, most of the well-studied vdW ferromagnets like CrX$_3$ (X $=$ Br, Cr, I) lack at least one of the desired attributes discussed above. On the other hand, the family of compounds with the general formula Fe$_n$GeTe$_2$ (with $n\geq$3) are potentially outstanding materials for spintronic applications as they show metallic conductivity, near-room temperature ferromagnetism and for being effectively 2D materials they can be easily integrated with other layers of functional materials through vdW nano-structuring\cite{Yang(a),Zhang}. Within this family, $n=$ 3 provides most efficient decoupling between the 2D layers, but shows a lower $T_{Curie}$. The compund with $n=5$ displays the highest $T_{Curie}$, but with relatively stronger coupling between the layers. Fe$_4$GeTe$_2$ ($n=$4) falls in between where $T_{Curie}$ is near room temperature, and the 2D layers are sufficiently decoupled for easy exfoliation \cite{Mondal,Seo,Bera}. Such properties make Fe$_4$GeTe$_2$ the most promising candidate for spintronic applications. In this paper we show that Fe$_4$GeTe$_2$ is capable of generating a highly spin-polarized transport current owing to a spin-split band structure along with a higher Fermi velocity of the majority spin channel.

 Fe$_4$GeTe$_2$ is a van der Waals layered material that crystallizes in a rhombohedral structure with the space group $R\bar{3}m$ \cite{Seo} as shown in the Figure 1(a). The Fe atoms occupy two inequivalent Wyckoff sites denoted as Fe-I and Fe-II.  It has a middle Ge layer with Fe-I and Fe-II dumbells placed alternately above and below its plane. Two hexagonal layers of Te encapsulate these  Fe$_4$Ge layers from top and bottom, forming  Fe$_4$GeTe$_2$ layer, and a van der Waals gap separates these layers. Transport and magnetization measurements reveal that Fe$_4$GeTe$_2$ is an itinerant ferromagnet with good metallic conductivity down to the cryogenic temperatures\cite{Bera}. High quality single crystals (structure shown in Figure 1(a)) of the material was grown by chemical vapor transport method which yielded reasonably large size (2 mm $\cross$ 2 mm) crystals. Thanks to the layered structure of the crystals, it was possible to expose a pristine surface by mechanical cleaving prior to the low temperature experiments. An atomic resolution image (captured at 77 K) of the Te-terminated surface of one of our crystals is shown in Figure 1(b). Such atomic resolution was obtained everywhere on the surface confirming the good surface quality. In Figure 1(c) we show a larger area image where a unit cell step of height 1.6 nm is seen. As shown in Figure 1 (d), the crystals show a ferromagnetic transition near 300 K. Further discussion on the temperature and magnetic field dependence of magnetization are discussed elsewhere\cite{Bera}. The low-temperature Andreev reflection experiments were done using a conventional needle-anvil technique using superconducting tips. A home-built point contact probe along with the tip and the crystal was mounted in a liquid helium based cryostat (American magnetics) equipped with a 6T-1T-1T vector magnet. The $dI/dV$ vs. $V$ spectra were recorded using a lock-in modulation technique at 882 Hz. The lock-in modulation technique is discussed in detail in Section III of the supplementary information. The schematic of tip and sample forming point contact is shown in Figure 1(e). 
\begin{figure}[h!]
 \centering
  \includegraphics[scale=0.35]{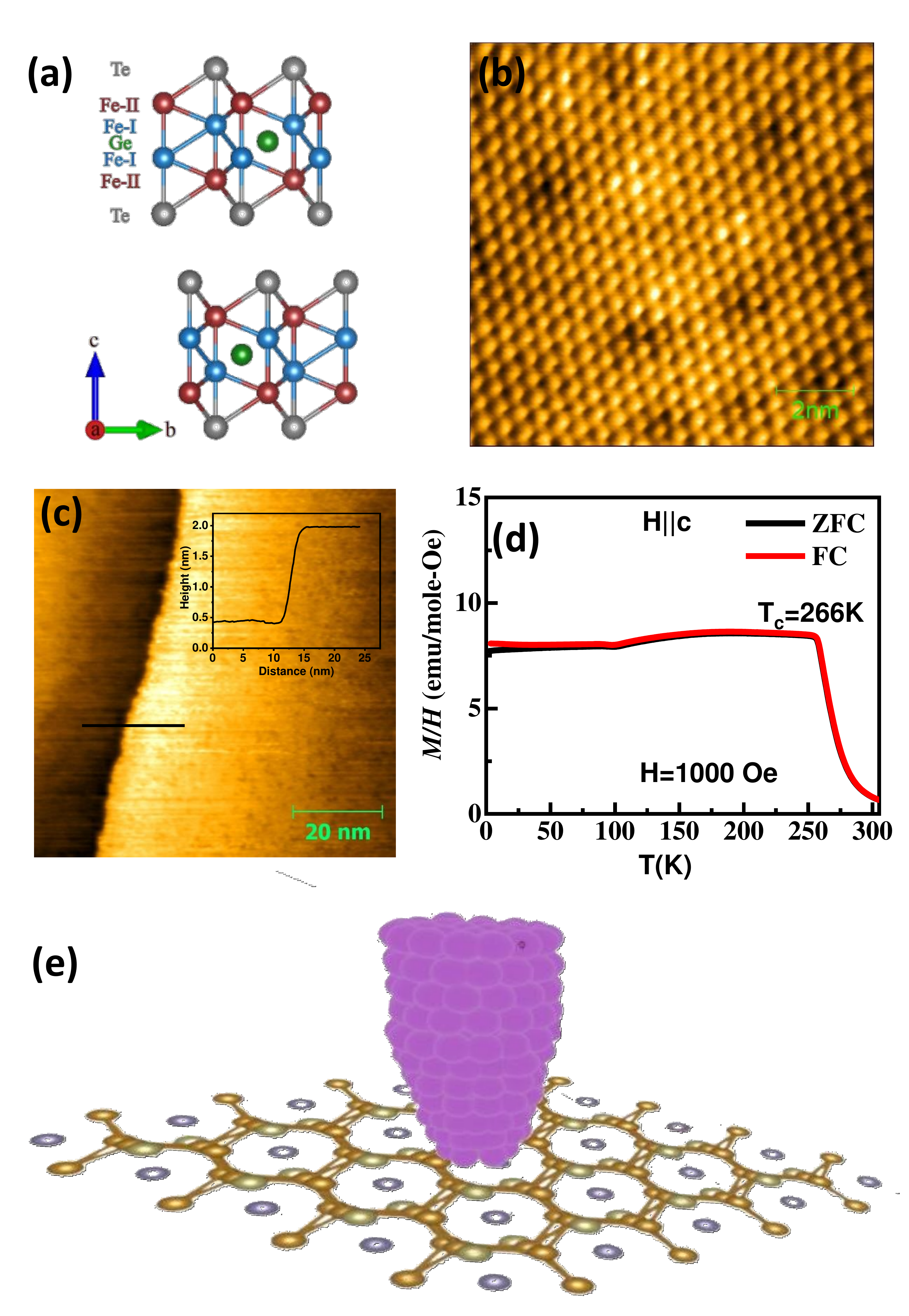}

	\caption{{(a)Crystal structure of Fe$_4$GeTe$_2$. (b) STM topograph of  10.43 nm $\times$ 10.43 nm area taken on the cleaved surface of Fe$_4$GeTe$_2$ at sample bias $V_b$= 0.82 V and tunneling current $I_t$ =200 pA. The atoms are clearly resolved. (c) STM topograph of area 80 nm  $\times$ 80 nm showing step taken at $V_b$= 0.9 V and tunneling current $I_t$ =120pA. The height of the step is shown in the inset. (d) Temperature dependence of the dc magnetisation of Fe$_4$GeTe$_2$ in zero field and field cooled state taken at $\mu{_0}H$ = 1000 Oe for H$\parallel$c. (e) Schematic showing a tip and sample forming point contact.}}
\end{figure}

\begin{figure}[h!]
 \centering
  \includegraphics[scale=0.35]{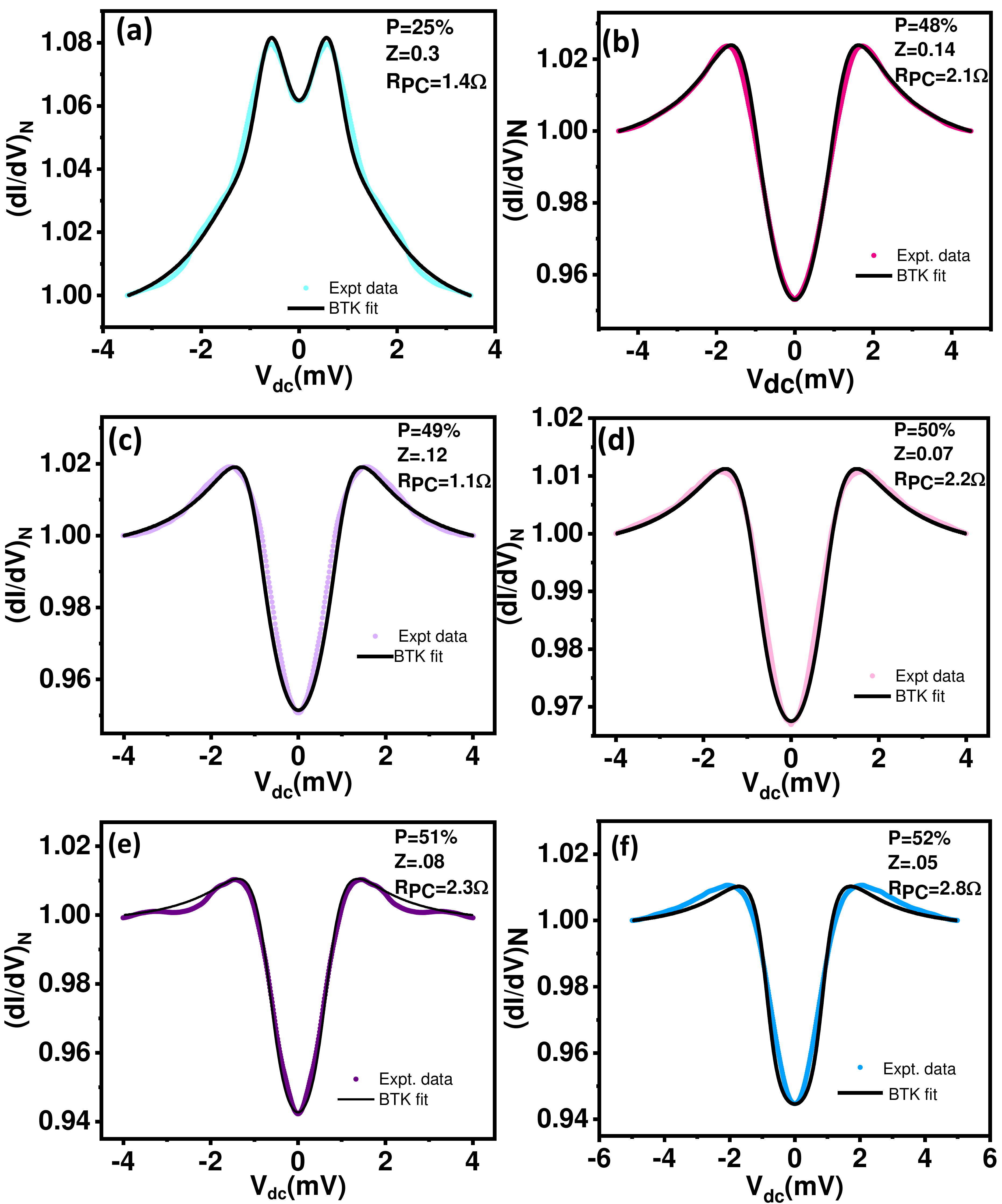}

	\caption{{(a)-(f) Six fitted ballistic spectra using Nb tip with different values of barrier strength (Z) and spin polarisation ($P_t$).
  }}
\end{figure}
In Figure 2, we show six representative normalized Andreev reflection spectra ($dI/dV$ vs. $V$) captured at 1.6 K on Fe$_4$GeTe$_2$ with superconducting Nb tips. One generic feature of all those spectra is a heavily suppressed Andreev reflection\cite{Blonder, Mazin}. Ideally, the conductance enhancement at zero-bias is expected to be 200\%, but in the Nb/Fe$_4$GeTe$_2$, the enhancement is only $\sim$ 2\%. Such dramatic suppression of Andreev reflection is linked to a highly spin-polarized Fermi surface of Fe$_4$GeTe$_2$ \cite{Soulen}. To extract the magnitude of spin polarization we first focus on the physics of charge transport at metal-superconductor interfaces. The transport through a metal-superconductor junction is dominated by Andreev reflection which involves a spin up (down) electron from the metal incident on the interface reflect as a spin down (up) hole. The Andreev reflection spectra ($dI/dV$ versus $V$) obtained through such junctions are usually analyzed in the framework of the theory given by Blonder, Tinkham, and Klapwijk (BTK)\cite{Blonder}. This theory employs a $\delta$-function potential barrier of the form $V(x)$=$V_0$ $\delta$(x) at the interface, whose strength can be characterized by a dimensionless quantity $Z = V_0/\hbar v_F$. In simple systems involving normal metals and conventional superconductors, the Andreev reflection spectra can be analyzed using two fitting parameters, $Z$ and $\Delta$, the superconducting energy gap. For more complex systems, often a broadening parameter ($\Gamma$) is needed and that originates from multiple mechanisms including finite quasipartcile lifetime, distribution of $\Delta$ \cite{Raychaudhuri(a)} and two-level fluctuations \cite{Wei}. When the metal in the metal-superconductor junction is a ferromagnet, that may host a spin-polarised Fermi surface --  the density of states of spin-up electrons ($N_\uparrow(E_F)$) is not equal to density of states of spin-down electrons ($N_\downarrow(E_F)$). In such a situation, all the electrons at the Fermi level cannot undergo Andreev reflection simply because all the corresponding Andreev reflected holes cannot find accesible states in the opposite spin band. This leads to a suppression of Andreev reflection.  By measuring the degree by which Andreev reflection is suppressed, the magnitude of spin-polarization at the Fermi surface
can be estimated\cite{Soulen}. This can be done using the following algorithm. (i) Calculate the BTK current for the unpolarized case $I_{BTKu}$, (ii) Calculate the BTK current for the fully polarized case spin polarization is 100 percent ($I_{BTKp}$) with all other fitting parameters including the Z remaining the same, (iii) Find an intermediate total polarization current by interpolation between $I_{BTKu}$ and $I_{BTKp}$ using  
\begin{equation}
I_{total}=I_{BTKu}(1-P_t)+I_{BTKp}P_t
\end{equation}
\begin{figure}[h!]
 \centering
  \includegraphics[scale=0.35]{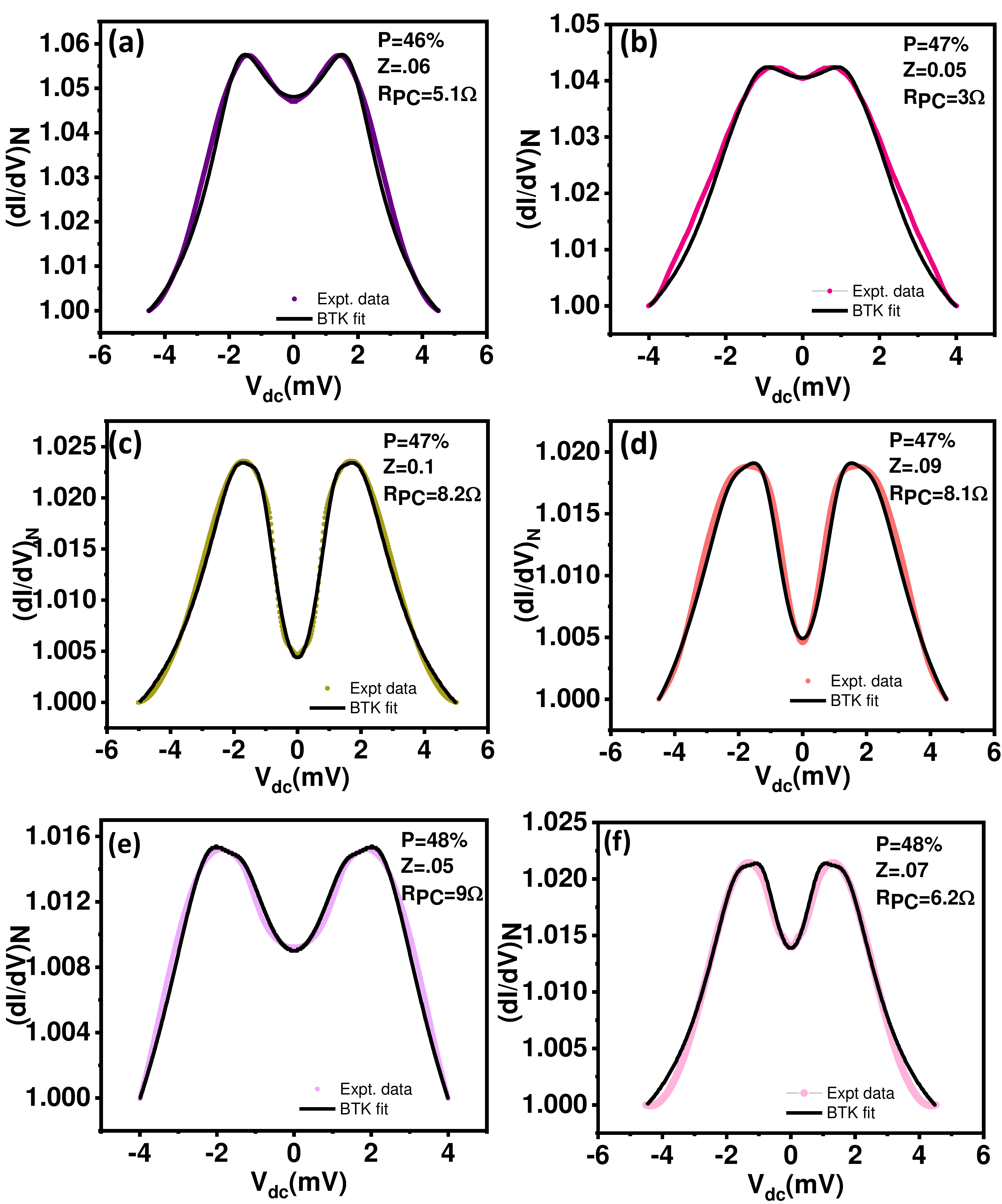}

	\caption{{(a)-(f) Six fitted ballistic spectra using Pb tip with different values of barrier strength (Z) and spin polarisation ($P_t$).
  }}
\end{figure}

Note that, the reason for keeping the parameters same for $I_{BTKu}$ and $I_{BTKp}$ is that these calculations are done for the same point contact. The derivative of I$_{total}$ with respect to $V$ gives the modified Andreev reflection spectrum
with finite spin polarization in the ferromagnet. Hence, the modified BTK  model\cite{Soulen} can be used
to analyze the spin-polarized Andreev reflection spectra obtained for ferromagnetic metal and a superconductor junctions, where the magnitude of $P_t$ is obtained as an important fitting parameter\cite{Jong, Upadhyay,Auth,Rana,Howlader,Kamboj,Aggarwal,Sirohi}. The details of the modified BTK model are discussed in Section IV of the supplementary information. We have followed the above algorithm to determine $P_t$ from the Nb/Fe$_4$GeTe$_2$ Andreev reflection spectra. The black lines over the colored experimental data points in Figure 1 shows the best fit to each individual spectrum with the modified BTK theory discussed above. Owing to the high surface quality of the crystal surface, we could establish ballistic (non-thermal) point contacts free from contact-heating related artifacts. The analysis (Figure 4(a)) reveals a very high degree of intrinsic spin polarization ($P_t$) approaching $\sim 51\%$.

As it is usually seen for Nb point contacts, $P_t$ decays with increasing barrier strength $Z$ possibly due to spin-flip scattering at less transparent barriers\cite{Sirohi}. In order to check the reproducibility of such a high value of $P_t$, we also performed Andreev reflection spectroscopy with Pb tips. The corresponding experimental spectra (color plots) and the modified BTK fits (black lines) are shown in Figure 3. Unlike for Nb/Fe$_4$GeTe$_2$ junctions, high $Z$ contact was not found for Pb/Fe$_4$GeTe$_2$ junctions. However, the analysis of the spectra (Figure 4(a)) for Pb/Fe$_4$GeTe$_2$  reveal $P_t \sim 49\%$ for low $Z$ contacts, thereby confirming that the measured high spin polarization is intrinsic to Fe$_4$GeTe$_2$. Here it should also be noted that due to the unavailability of a high temperature elemental superconductor, our measurements were performed  below the critical temperature of Nb ($T_c \sim$ 9 K) and Pb ($T_c\sim$7.2 K). It was earlier shown that the transport spin polarization of itinerant ferromagnets vary with temperature in the same way as the bulk magnetization \cite{Sourin}. As it can be seen from Figure 1(d), the bulk magnetization does not vary noticeably  near room temperature confirming that the transport spin polarization ($P_t$) of Fe$_4$GeTe$_2$ is approximately 50\% up to such temperatures. The value of fitting parameters used in the analysis of spectra shown in Figure 2 and 3 can be found in Section V of the Supplementary information.

Now it is important to understand the origin of such a large spin polarization. To shed light on that, we calculated the degree of transport spin polarization of the van der Waals ferromagnet Fe$_4$GeTe$_2$ within the first-principles density functional theory as implemented in the Vienna Ab-initio Simulation Package (VASP)\cite{Hohenberg, Kohn, Kresse, Kresse(A)}. The wavefunctions are described within the projector augmented wave formalism \cite{Bloch}, with \SI{700}{\electronvolt} cutoff for the kinetic energy. The local density approximation~\cite{Perdew} is used to describe the exchange-correlation energy and non-local optB86b correlation functional to account for the interlayer van der Waals interactions\cite{Klime}. We sampled the first Brillouin zone with $17\times 17\times 5$ Monkhorst-Pack $k$–mesh \cite{Monkhorst}and optimized the crystal structure until the forces were below \SI{1}{\milli \electronvolt \per \angstrom}.

\begin{figure}[h!]
 \centering
  \includegraphics[scale=0.3]{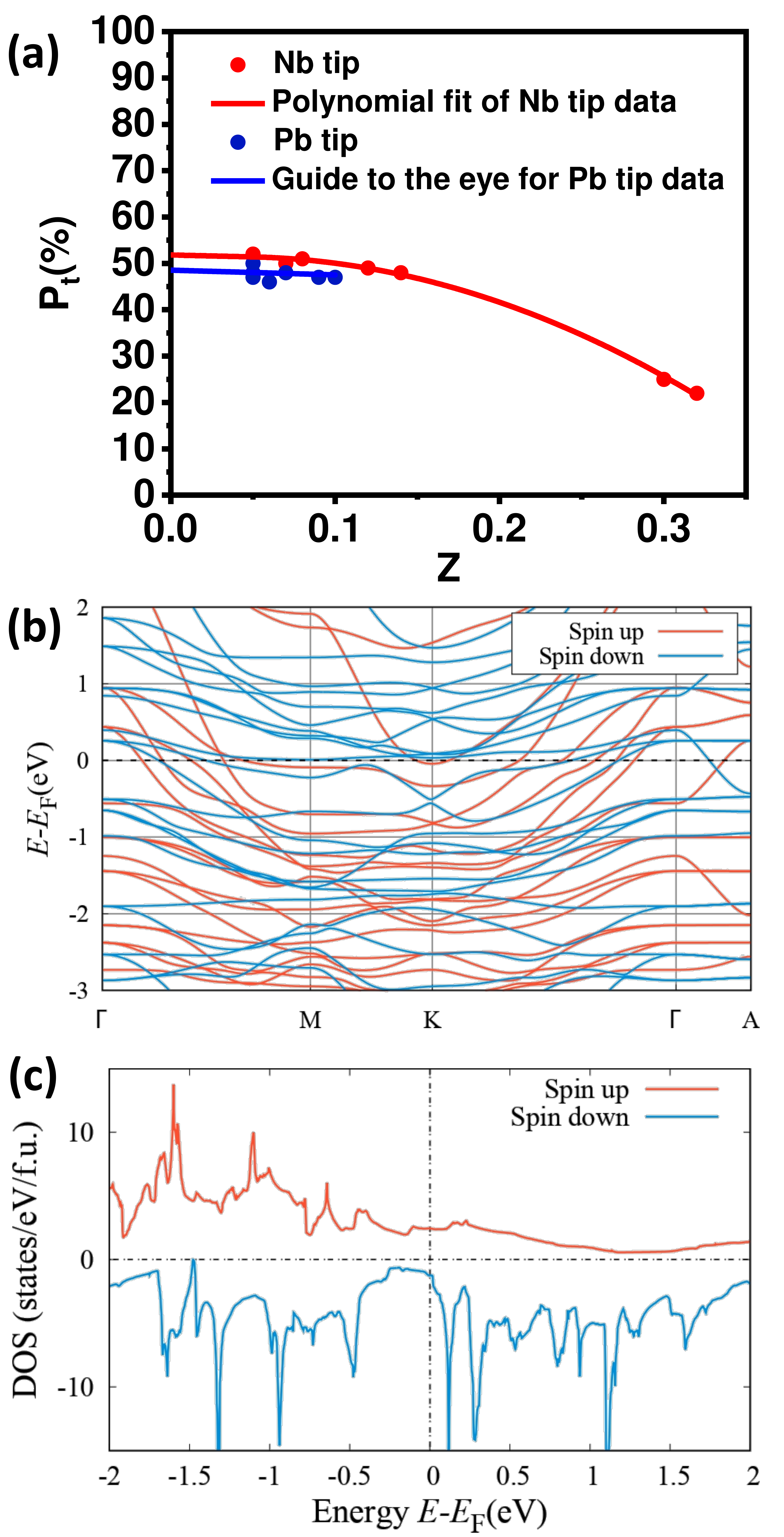}

	\caption{{ $P_t$ vs Z for point contact junctions of Fe$_{4}$GeTe$_2$ with Nb and Pb tips.   (b) Spin-polarized band structure and (c) density of states calculated for the ferromagnetic
    state.
  }}
\end{figure}

The optimized lattice parameters $a = \SI{3.94}{\angstrom}$ and $c = \SI{28.87}{\angstrom}$ agree with the experimental results\cite{Seo}. Fe-I is covalently bonded to Ge and has a magnetic moment of 1.66 $\SI{} {\mu_B}$, considerably lower than Fe-II (2.47 $\SI{} {\mu_B}$). The weighted average of 2.06 $\SI{}{\mu_B \per Fe}$ corresponds well with the experimental value of 1.8 $\SI{}{\mu_B \per Fe}$ \cite{Seo,Bera}. The spin-polarized electronic structure analysis and density of states [Figure 4(b)-(c)] indicate metallic nature, which along with the non-integer magnetic moments, imply the itinerant ferromagnetism. 
The electronic structure shows several flat bands near the Fermi level, indicating enhanced effective mass, suggesting a strong electron correlation. The states of the majority spin channels near the Fermi level are composed of $d_{yz}$, $d_{z^2}$, and $d_{xz}$ orbitals of Fe-I and $d_{z^2}$ of Fe-II hybridized with $p$ orbitals of Te. Whereas $d_{xy}$ and $d_{x^2-y^2}$ of Fe-I and $d_{z^2}$ of Fe-II constitute the states for the minority spin channel.

Since Andreev reflection spectroscopy measurements are essentially transport measurements, they are sensitive to the spin polarization at the Fermi energy and not that of the whole spin-polarized bands. In this context, the more relevant quantity is transport spin polarization which essentially gives a direct estimate of the spin-polarized current (the transport spin polarization), a truly relevant quantity for spintronic applications. Within the classical Bloch-Boltzman transport theory, the transport spin polarization in terms of spin-dependent currents takes the form{\cite{Mazin}, 
\begin{equation}
P_t^n = \frac{\langle N(E_{\rm F})v^n_{\rm F}\rangle_{\uparrow} - \langle N(E_{\rm F})v^n_{\rm F}\rangle_{\downarrow}}{\langle N(E_{\rm F})v^n_{\rm F}\rangle_{\uparrow} + \langle N(E_{\rm F})v^n_{\rm F}\rangle_{\downarrow}},  \nonumber 
\end{equation}  
where $v_{\rm F}$ is the spin-polarized Fermi velocity. Calculated $P_t^n$ corresponds to the spin-polarization without considering the contribution of the Fermi velocity mismatch between the up and down bands ($(n = 0)$) and the transport spin polarization in the ballistic $(n = 1)$ and diffusive $(n = 2)$ regimes of transport.  We found the average Fermi velocity of the majority spin channels to be $v_{\rm F \uparrow} = \SI{4.37 E5}{\meter \per \second}$ and a considerably lower value for minority spin channels, $v_{\rm F \downarrow} = \SI{1.88 E5}{\meter \per \second}$. The calculated spin-polarized density of states at the Fermi level for the majority carriers is $N_{\uparrow} = \SI{2.47}{states \per \electronvolt \per f.u.}$, which is 96\% greater than the $N_{\downarrow}$. This significant imbalance in the Fermi velocities and density of states of the two channels results in a large value of transport spin polarization ($P_t$) of 64\% in ballistic and 82\% in the diffusive regimes. The comparison of the calculated $P_t$ with our experimental results further confirm that our point contact measurements provided a value near the ballistic limit which is consistent with the ballistic spectral features that we obtained from experiments (Figure 2 \& 3). The minor difference ($\sim 8\%)$ can be attributed to the fact that there is no way that the role of the parameter $Z$ can be included in the calculations. A finite $Z$ is unavoidable in experiments and that gives rise to a lower estimate of $P_t$. In addition, the broad non-linear background of the spectra may include an error in the estimate \cite{Rana}. The origin of discrepancy can also be due to the presence of flat bands near the Fermi energy that can lead to correlation effects and hence affect the estimate of transport spin polarisation. After considering all the factors, the measured spin polarization exceeding 50\% for Fe$_4$GeTe$_2$ makes it an extremely important candidate for spintronic applications.

In conclusion, we have performed spin-resolved Andreev reflection spectroscopy on the van der Waals ferromagnet Fe$_4$GeTe$_2$ using superconducting Nb and Pb tips. The measurements reveal a very high degree of transport spin polarization exceeding 50\% in Fe$_4$GeTe$_2$. The high transport spin polarization along with good metallic character and high Curie temperature of Fe$_4$GeTe$_2$ makes the system an important candidate for application in low-power spintronic devices.

 GS acknowledges Science and Engineering Research
Board, Government of India, for the CRG/2021/006395 Core Research Grant. DR and MB acknowledge the Department of Science and Technology, Government of India for the INSPIRE Fellowship. BG and MK acknowledge the support and resources provided by the PARAM Brahma Facility at the Indian Institute of Science Education and Research, Pune, under the National Supercomputing Mission of the Government of India. SB acknowledges CSIR Grant No. 09/080(1110)/2019-EMR-I. MM acknowledges the Department of Science and Technology, Government of India (Grant No. SRG/2019/000674).


\end{document}


\title{Supplementary information  \\
High transport spin polarization in the van der Waals ferromagnet Fe$_4$GeTe$_2$ }

\author{Deepti Rana$^1$, Monika Bhakar$^1$, Basavaraja G.$^2$, Satyabrata Bera$^3$, Neeraj Saini$^1$, Suman Kalyan Pradhan$^3$, Mintu Mondal$^3$, Mukul Kabir$^2$ and Goutam Sheet$^{1}$} 

\email{goutam@iisermohali.ac.in}

\affiliation{$^1$Department of Physical Sciences, Indian Institute of Science Education and Research (IISER) Mohali, Sector 81, S. A. S. Nagar, Manauli, PO 140306, India}

\affiliation{$^2$Department of Physics, Indian Institute of Science Education and Research (IISER) Pune, PO 411008, India}

\affiliation{$^3$School of Physical Sciences, Indian Association for the Cultivation Of Science, Jadavpur, Kolkata-700032}

\maketitle
	\section{X-ray diffraction} 
	The X-ray diffraction pattern, shown in Figure S1,  of the cleaved surface of  Fe4GeTe2  single crystal taken at room temperature. All the Bragg reflection peaks come from (00l) plane. 
\begin{figure}[!h]
\centering
\includegraphics*[width=0.8\linewidth]{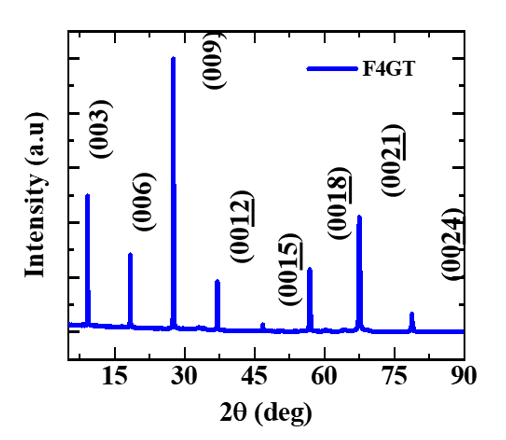}
\caption{ X-ray diffraction pattern of Fe$_4$GeTe$_2$.}
\end{figure}

\section{Fast Fourier Transform} 
	Figure S2 shows the fast Fourier transform (FFT) of the atomically resolved STM topography (shown in Figure 1(b)). The six-fold symmetry of the crystal can be seen.
	
\begin{figure}[!h]
\centering
\includegraphics*[width=0.65\linewidth]{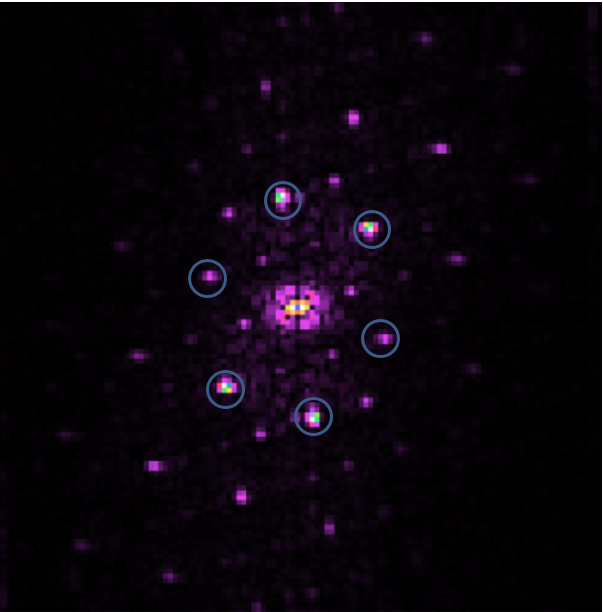}
\caption{ FFT of STM topograph in Figure 1(b).}
\end{figure}

	\section{Experimental setup and data acquisition}
    Point contact spectroscopy is a standard technique for the measurement of spin polarization of a metallic ferromagnet. There are numerous ways in which a point contact junction can be formed between two different materials. One of the commonly used methods is the ‘needle-anvil method’ in which a sharp tip is approached, either using a piezo positioner or a differential screw, towards the sample until a light ‘ohmic’ contact is made. We have used a differential screw, having 100 threads per inch, which can be operated from the top of the probe using a manipulator. After the point contact between the tip and the sample is established, a lock-in-based modulation technique to acquire point contact spectrum ($dI/dV$ $vs$ $V$). A schematic of the setup is shown in Figure S3. In this technique, a sweeping DC current ($I_{dc}$) is coupled to fixed low-frequency ac current (in the range of a few Hertz to a few kilo-Hertz). The dc current source (Kethley 6220) is used to supply dc current. The lock-in amplifier (Stanford Research System  830) can be used to provide ac current by using voltage to current converter. The dc voltage drop is then recorded using a digital multimeter (Kethley 2000). The  ac voltage output of the lock-in amplifier, for signal locked at the first harmonic,  is proportional to  differential resistance dV/dI from which differential conductance is calculated. 

\begin{figure}[!h]
\centering
\includegraphics*[width=0.8\linewidth]{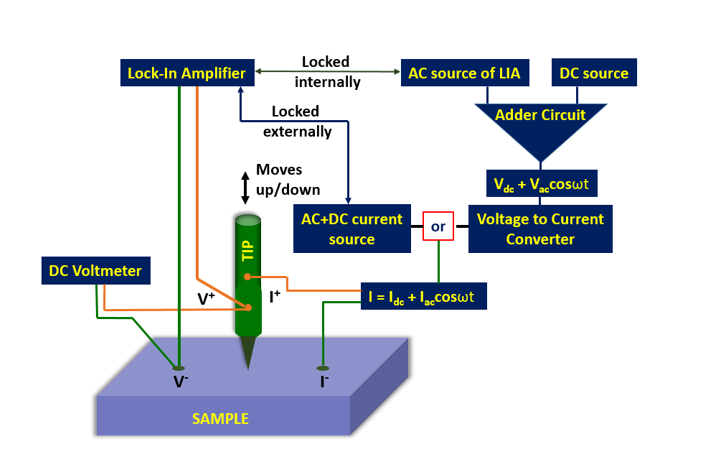}
\caption{ Schematic showing the point contact spectroscopy measurement setup.}.
\end{figure}

\section{Modified BTK theory}
The current for point contact geometry between a normal metal and superconductor, as per BTK theory\cite{Blonder}, is given by:
\begin{equation}
    I_{BTKu}=C \int_{-\infty}^{\infty} [f(E-eV)-f(E)][1+A_{BTKu}(E)-B_{BTKu}(E)] dE
\end{equation}

where $A_{BTKu}(E)$ and $B_{BTKu}(E)$ correspond to probabilities of Andreev reflection and normal reflection respectively for the case of unpolarised current.

For the fully polarised cuurent $I_{BTKp}$, equation 1 can be used by modifying $A_{BTKu}$ and $B_{BTKu}$ to  $A_{BTKp}$ and $B_{BTKp}$ where the latter are the corresponding probablities for the fully polarised case.

The coefficients $A_{BTKu}(E)$, $B_{BTKu}(E)$, $A_{BTKp}(E)$ and $A_{BTKp}(E)$ can be  calculated in the following way:\\
$A_{BTKp}(E)$=0 for all values of $E$\\
$B_{BTKp}(E)$=1 for all values of $E<\Delta$\\
$B_{BTKp}(E)$=$\frac{(\sqrt{\frac{E^{2}-\Delta^{2}}{E^2}}-1)^{2}+4Z^{2}(\frac{E^{2}-\Delta^{2}}{E^2})}{(\sqrt{\frac{E^{2}-\Delta^{2}}{E^2}}+1)^{2}+4Z^{2}(\frac{E^{2}-\Delta^{2}}{E^2})}$ for $E>\Delta$\\
$A_{BTKu}(E)$=$\frac{\Delta^{2}}{E^{2}+(\Delta^{2}-E^{2})(1+2Z^{2})^{2}}$ for $E<\Delta$\\
$A_{BTKu}(E)$=$\frac{(u^{sc}v^{sc})^2}{\gamma^2}$ for $E>\Delta$\\
$B_{BTKu}(E)$=$1-A(E)$ for $E>\Delta$
$B_{BTKu}(E)$=$\frac{((u^{sc})^{2}-(v^{cs})^2)^{2}Z^{2}(1+Z^{2})}{\gamma^2}$for $E>\Delta$\\
$\gamma^{2}=(((u^sc)2-(v^sc))^{2}Z^{2}+(u^{sc})^{2})^{2}$

where $u^{sc}$ and $v^{sc}$ comes from the solution of Bogoliubov-de Gennes (BdG) equation for  superconductor as :
$(u^sc)^2$=$1-(v^sc)^2$=$\frac{1}{2}(\sqrt{\frac{E^{2}-\Delta^{2}}{E^{2}}})$

The total current in terms of $I_{BTKu}$ and $I_{BTKp}$ can then be written as :

\begin{equation}
    I_{total}= (1-P_{t}) I_{BTKu}+P_{t} I_{BTKp}
\end{equation}
where $P_t$ is one of the fitting parameter of the modified BTK theory which gives an estimate of the value of tranport spin polarisation \cite{Mazin, Soulen}.

\begin{figure}[!h]
\centering
\includegraphics*[width=1\linewidth]{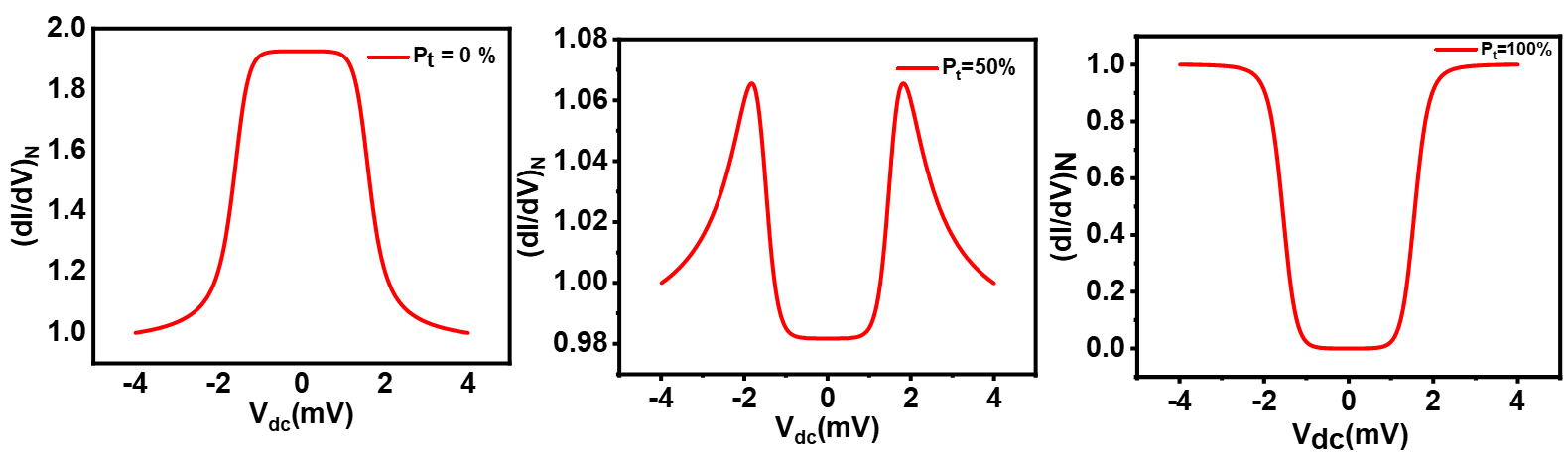}
\caption{ Three representative ballistic spectra for (a) unpolarised case($P_t$=0 \%), (b) partially polarised case ($P_t$=50\%) and fully polarised case ($P_t$=100\%).}.
\end{figure}

Three representative $dI/dV$ spectra have been shown in Figure S4 for the case when one of the system making the point contact junction has (a) $P_t$ =0, (b) $P_t$ =50\%  or (c) $P_t$ =100\%. It is to be noted that these spectra were plotted for values of $T$ = 1.7K, $Z$=0, $\Delta$ =1.5meV and $\Gamma$=0 meV.

\section{{Fitting parameters}}

Table I and II contains the list of fitting parameters used for fitting of spectra obtained for Nb/Fe$_4$GeTe$_2$ (Figure 2) and Pb/Fe$_4$GeTe$_2$ (Figure 3) respectively.

\begin{table}[!h]
\centering
\includegraphics*[width=0.8\linewidth]{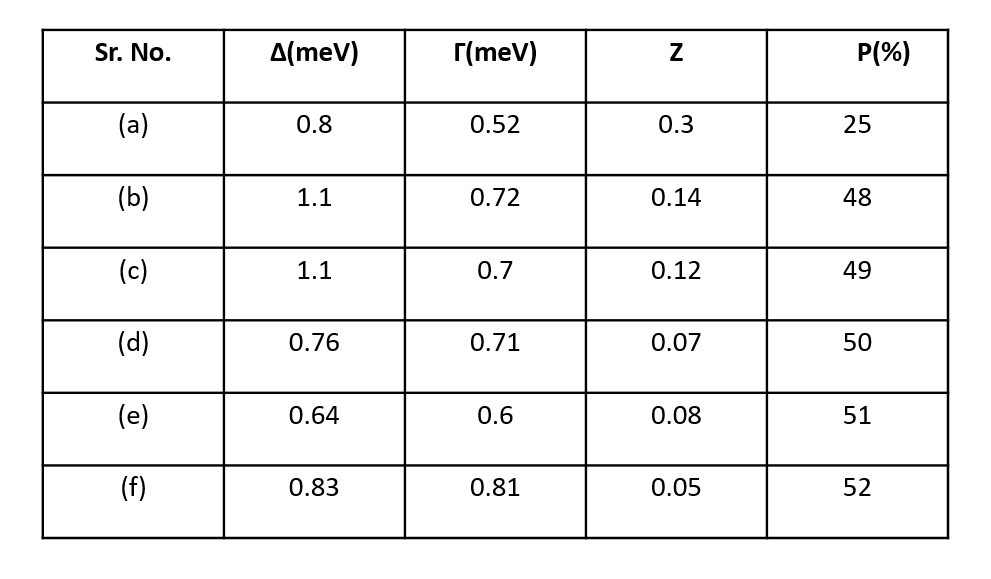}
\caption{List of fitting parameters for Figure 2 of the main manuscript representing ballistic spectra for Nb/Fe$_4$GeTe$_2$ point contact.}
\end{table}
\begin{table}[!h]
\centering
\includegraphics*[width=0.8\linewidth]{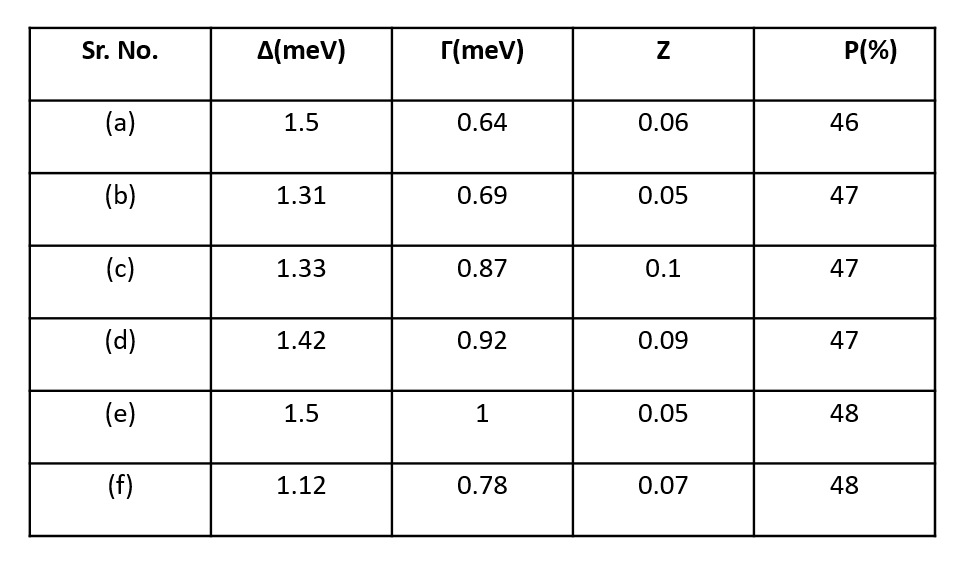}
\caption{List of fitting parameters for Figure 3 of the main manuscript representing ballistic spectra for Pb/Fe$_4$GeTe$_2$ point contact.}
\end{table}

\pagebreak